\documentclass[twocolumn]{aastex631}
\usepackage{CJK}

\newcommand{\kms}{{km\,s$^{-1}$}\xspace} 

\newcommand{\ergs}{{erg\,s$^{-1}$}\xspace}

\usepackage{xspace}
\usepackage{color}
\makeatletter
\DeclareRobustCommand{\erase}{\bgroup\markoverwith{\textcolor{red}{\rule[.5ex]{2pt}{1.5pt}}}\ULon}

\makeatother

\submitjournal{Astrophysical Journal}

\begin{document}
\begin{CJK*}{UTF8}{gbsn}
\title{
    VLBI Detection of an Active Radio Source Potentially Driving 100-kpc Scale Emission in the Ultraluminous Infrared Galaxy IRAS\,F01004$-$2237
}



\correspondingauthor{Takayuki J. Hayashi}
\email{t.hayashi@nao.ac.jp, t.hayashi.bak@gmail.com}

\author[0000-0002-4884-3600]{Takayuki J. Hayashi}
\affiliation{National Astronomical Observatory of Japan \\
2-21-1 Osawa, Mitaka \\
Tokyo 181-8588, Japan}
\affiliation{Azabu Junior and Senior High School\\
2-3-29 Motoazabu, Minato\\
Tokyo 106-0046, Japan}

\author[0000-0002-9043-6048]{Yoshiaki Hagiwara}
\affiliation{Toyo University \\
5-28-20 Hakusan, Bunkyo \\
 Tokyo 112-8606, Japan}

\author[0000-0001-6186-8792]{Masatoshi Imanishi}
\affiliation{National Astronomical Observatory of Japan \\
2-21-1 Osawa, Mitaka \\
Tokyo 181-8588, Japan}
\affiliation{Toyo University \\
5-28-20 Hakusan, Bunkyo \\
 Tokyo 112-8606, Japan}
\affiliation{Department of Astronomy, School of Science, Graduate University
for Advanced Studies (SOKENDAI)\\
2-21-1 Osawa, Mitaka \\
Tokyo 181-8588, Japan}

\begin{abstract}
The nearby ultraluminous infrared galaxy (ULIRG) IRAS\,F01004$-$2237 exhibits 100-kpc scale continuum emission at radio wavelengths. 
The absence of extended X-ray emission in IRAS\,F01004$-$2237 has suggested an active galactic nucleus (AGN) origin for the extended radio emission, whose properties and role in merging systems still need to be better understood. 
We present the results of multi-frequency observations of IRAS\,F01004$-$2237 conducted by the Very Long Baseline Array at 2.3 and 8.4\,GHz.
Compact 8.4-GHz continuum emission was detected on a 1-pc scale in the nuclear region with an intrinsic brightness temperature of $10^{8.1}$\,K suggesting that the radio source is originated from an AGN, potentially driving the extended emission.
In contrast, no significant emission was observed at 2.3\,GHz, indicating the presence of low-frequency absorption.
This absorption cannot be attributed solely to synchrotron self-absorption; alternatively, free-free absorption due to thermal plasma is mainly at work in the spectrum.
From combined perspectives, including mid-infrared and X-ray data, the AGN is obscured in a dense environment.
The kinetic power of the nonthermal jet, as inferred from the extended emission, can play a more important role in dispersing the surrounding medium than the thermal outflow in IRAS\,F01004$-$2237.
These findings hint that jet activities in ULIRGs may contribute to  AGN feedback during galaxy evolution induced by merger events.
\end{abstract}


\section{Introduction} \label{sec:intro}
Galaxies with large infrared luminosity greater than $10^{12}L_{\odot}$ are known as ultraluminous infrared galaxies \citep[ULIRGs;][]{1996ARA&A..34..749S}. 
ULIRGs are systems of a major merger and harbor substantial quantities of infrared-emitting dust centrally concentrated within their nuclei \citep{1988ApJ...325...74S,1996MNRAS.279..477C,2000ApJ...529L..77B,2002ApJS..143..315V}. 
Their powerful energy sources are energetic radiation from active galactic nuclei (AGNs) and starbursts 
\citep{1998ApJ...498..579G,2009ApJS..182..628V}.
Both phenomena are activated by the inflow of gas resulting from merger events, with the former possessing the capacity to terminate the latter by expelling the gas
\citep{2006ApJS..163....1H,2008ApJS..175..356H}.
The role of AGNs during mergers remains a subject of contention, closely linked to understanding the history of star formation and the growth of supermassive black holes within the obscured galactic population during the early Universe \citep{2006asup.book..285L,2021A&ARv..29....2P}.

When AGN activities halt the preceding star formation (AGN feedback), outflows emanating from the central engine are inferred to play a crucial role \citep[e.g.,][]{1998A&A...331L...1S,2012A&A...537L...8C,2015ApJ...799...82C,2016A&A...591A..28C,2024MNRAS.52710844H}, where non-thermal jets are also supposed to contribute to this phenomenon \citep[e.g.,][]{2017ApJ...844...37D,2017A&A...599A.123N,2023ApJ...959..107S,2023MNRAS.520.5712S}.
Despite its importance, many questions about radio emission from AGNs in ULIRGs still need to be answered. 
A well-established correlation, spanning up to five orders of magnitude, exists between far-infrared and radio emission from galaxies \citep[e.g.,][]{1985ApJ...298L...7H,1991ApJ...376...95C,1991ApJ...378...65C,2004ApJS..154..147A,2019ApJ...875...80G}.
The intense radio emission, deviating significantly from this correlation, is attributed to the influence of AGN activities \citep[e.g.,][]{2001ApJ...554..803Y,2017MNRAS.469.3468C,2024arXiv240104924W}.
However, recent studies have shown that radio excess can only  collect the most radio-loud AGNs \citep[e.g.,][]{2010ApJ...724..779M,2019PASJ...71...28S}.
Investigating the individual AGN activity at radio wavelengths remains crucial, even for objects whose radio excess has not been detected \citep[cf.][]{2012ApJS..203....9U}.

This work will focus on the nearby ULIRG IRAS\,F01004$-$2237,  located at $z=0.118$, with an infrared luminosity of $10^{12.2} L_{\odot}$ \citep{1998ApJS..119...41K}. 
The object is categorized as an old merger, displaying slight distortions similar to those observed in mergers but lacking direct indications of tidal tails at optical wavelengths \citep{2002ApJS..143..315V}.
While this object does not exhibit a pronounced radio excess, aligning with the far-infrared-radio correlation,
it has been classified as a type-2 Seyfert based on optical observations \citep{1991MNRAS.248..528A,2010ApJ...709..884Y,2013MNRAS.432..138R}.
Furthermore, \cite{2021MNRAS.504.2675H} have reported extended radio emission on a 100-kpc scale whose morphology is reminiscent  of radio galaxies.
Recently, large-scale ($\sim100$\,kpc) radio continuum emission has been detected in a select number of ULIRGs \citep[e.g.,][]{2021MNRAS.503.5746N,2022A&A...664A..25K}, and the properties of such emission in merging systems are still less known.
Observational studies of IRAS\,F01004$-$2237 will yield valuable insights into the role of AGN jets in ULIRGs.
At this moment, \cite{2021MNRAS.504.2675H} have suggested that no detection of extended X-ray emission \citep{1999A&A...349..389V,2005ApJ...633..664T} supports an AGN origin of the extended radio emission, not a radio mini-halo or a radio relic associated with galaxy clusters \citep{2012A&ARv..20...54F,2019SSRv..215...16V}.
However, because this argument is only circumstantial and eliminative, detecting a high-brightness radio emission from an AGN through high resolution observations at the milliarcsecond (mas) scale is necessary.

This paper presents the results of multi-frequency radio observations conducted  using the Very Long Baseline Array (VLBA). 
The high resolution achieved by very long baseline interferometry (VLBI) offers valuable insights into the presence of an AGN core responsible for the extended radio emission in IRAS\,F01004$-$2237.
While VLBI imaging observations of ULIRGs have been widely conducted, they have been confined to a limited number of objects. 
For instance, in Mrk\,231, a lobe structure and a recently emanating jet due to AGN activities have been identified \citep{1998AJ....115..928C,1999ApJ...519..185T,1999ApJ...516..127U,1999ApJ...517L..81U,2003ApJ...592..804L,2009ApJ...706..851R,2013ApJ...776L..21R,2017ApJ...836..155R,2020ApJ...891...59R,2021MNRAS.504.3823W}. 
A young radio source, indicative of the possible emergence of a new quasar in the context of merging galaxies, has been reported for IRAS\,00182$-$7112 \citep{2012MNRAS.422.1453N}.
In contrast, VLBI observations have also revealed radio supernovae (RSNe) and supernova remnants (SNRs), providing evidence of starburst activities in objects like Arp\,220 \citep{1998ApJ...493L..13L,1998ApJ...493L..17S,2005MNRAS.359..827R,2006ApJ...647..185L,2007ApJ...659..314P,2011ApJ...740...95B,2012A&A...542L..24B,2019A&A...623A.173V} and IRAS\,17208$-$0014 \citep{2003ApJ...587..160M,2006ApJ...653.1172M}.
Furthermore, composites of both AGNs and starbursts have been identified in several objects, such as IRAS\,23365+3604 \citep{2012MNRAS.422..510R} and Mrk\,273 \citep{2000ApJ...532L..95C,2005MNRAS.361..748B}. 
The new VLBA observations for IRAS\,F01004$-$2237 presented in this paper are expected to provide new insights into the origin of radio emission in the nuclear region of ULIRGs.

In this research, we employed the standard cosmological model with cold dark matter and a cosmological constant, adopting $H_0 = 70$\,\kms\,Mpc$^{-1}$, $\Omega_{\rm M} = 0.3$, and $\Omega_\Lambda = 0.7$, supported by observational studies from the past decades \cite[e.g.,][]{2020A&A...641A...6P}.
Under this model, at a redshift of the source ($z=0.118$), 1\,mas corresponds to a scale of 2.132\,pc, and the luminosity distance, $D_{\mathrm L}$, is 549.7\,Mpc.
The paper defines a spectral index, $\alpha$, as $S_\nu \propto \nu^\alpha$, where $S_\nu$ represents flux density at frequency, $\nu$. 

\section{Observations} \label{sec:obs}
The VLBA observation targeting IRAS\,F01004$-$2237 was executed on 2022 May 6, operating under the project code BH237C.
The observation comprised a 6-hour session involving all ten VLBA stations.
Four 128-MHz frequency subbands were employed throughout the observation for both right and left circular polarizations.
Simultaneous observations were conducted at the $S$ and $X$ bands. 
The $S$-band observation operated at a center frequency of 2.268\,GHz (hereafter 2.3\,GHz) with a bandwidth of 128\,MHz. 
The $X$-band observation spanned center frequencies of 8.304, 8.432, and 8.560\,GHz for each subband, with the central frequency being 8.432\,GHz (hereafter 8.4\,GHz) and a total bandwidth of 384\,MHz.
Data were sampled at the Nyquist rate with two bits per sample, resulting in a total data rate of 4\,Gbit\,s$^{-1}$. 
Only the parallel polarization hands were processed on the DiFX correlator  \citep{2011PASP..123..275D}.
Each 128-MHz subband was subdivided into 256 spectral channels to maintain wide-field imaging capabilities and deal with radio frequency interference.

The observations were conducted in phase-referencing mode. 
Scans lasting 3--4 minutes on IRAS\,F01004$-$2237 were interspersed with 2-minute scans on a calibrator, J0108$-$2157, positioned 1.3 degrees from IRAS\,F01004$-$2237 (R.A. = 01$^{\mathrm h}$\,08$^{\mathrm m}$\,19$^{\mathrm s}$.854172 and DEC. = $-$21$^\circ$\,57$'$\,38.75544$''$).
During the scans for the target, both the antennas and the correlator were aligned with the optical position of IRAS\,F01004$-$2237 (R.A. = 01$^{\mathrm h}$\,02$^{\mathrm m}$\,49$^{\mathrm s}$.920 and DEC. = $-$22$^\circ$\,21$'$\,57.000$''$).
The total integration time for the target amounted to 145\,minutes.
Table~\ref{tbl:obssum} provides a summary of the observations for IRAS\,F01004$-$2237.

\begin{deluxetable*}{ccccccccccc}
\tabletypesize{\scriptsize}
\tablewidth{0pt} 
\tablecaption{VLBA Observations Summary for IRAS\,F01004$-$2237. \label{tbl:obssum}}
\tablehead{								
	 \colhead{}	&	\colhead{}	&	\colhead{}	&	\colhead{}	&	\colhead{}	&	\colhead{}    &   \multicolumn{2}{c}{Beam Size}	\\\cline{7-8}
	 \colhead{Date} 	&	\colhead{Integration Time} &    \colhead{Phase Calibrator}  &   \colhead{Band}	&	\colhead{Frequency} &   \colhead{Band Width}    &      \colhead{FWHM} 	&	\colhead{PA} \\
	 \colhead{} 	&	\colhead{(min)}  &	\colhead{} &   \colhead{}	&	\colhead{(GHz)} 	&	\colhead{(MHz)} 	&	\colhead{(mas$^2$)} 	&	\colhead{(deg)} 	&	 	
	 } 							
\startdata 								
	2022 May 6 &	145 &    J0108$-$2157	&   $S$   &   2.268	&	 128	&	$7.01 \times 2.27$	&	$-1.6$	&	\\
	 &  & 	&    $X$   &	 8.432	&	 384	&	$1.75 \times 0.59$	&	\phs$0.1$		\\
\enddata												
\tablecomments{Simultaneous observations were conducted at both the $S$- and $X$-bands.  Beam sizes are determined in the uniformly weighted images.}
\end{deluxetable*}

\section{Data Reduction}
The data reduction followed a standard procedure using the Astronomical Image Processing System \citep[AIPS;][]{2003ASSL..285..109G} software developed at the National Radio Astronomy Observatory (NRAO). 
All analyses were performed independently in the 2.3-GHz and 8.4-GHz bands.
First, we flagged the data observed at an elevation angle of less than 15\,degrees to avoid large residuals from atmospheric propagation.
For the 8.4-GHz data, we also flagged the baselines with the Fort Davis station (FD), where some observational problems had been reported in the correlation processing. 
Following the digital sampler bias corrections with the task ACCOR, we performed an \textit{a priori} amplitude calibration with the task APCAL using system noise temperature measurements throughout the observation and gains provided by each station. 
Atmospheric opacity correction was not made in this step. 
Then, we corrected Earth orientation parameters and ionospheric dispersive delay on the 2.3-GHz and 8.4-GHz data using the tasks VLBAEOPS and VLBATECR, respectively.
Subsequently, fringe fitting of calibrators was performed on the data integrated over each scan using the task FRING. 
For the 8.4-GHz data, delays between subbands were also solved by the task MBDLY. 
Finally, we applied bandpass correction using the task BPASS for amplitude and phase using a fringe finder, S5\,0016$+$73. 
We confirmed that all calibrators were detected on all baselines at all frequencies, except for the baselines involving FD at 8.4\,GHz. 
At each step for the 2.3-GHz data, we identified and flagged radio frequency interference in the time and frequency domains using the tasks RFLAG and SPLAG.


Imaging processes were conducted using the CLEAN algorithm through the difmap software \citep{1997ASPC..125...77S}. 
First, we integrated the visibilities of the phase calibrator, J0108$-$2157, every 20\,seconds. 
Then, iterations of CLEAN and phase-only self-calibration were run by gradually reducing the solution time interval to 120, 60, and 30\,minutes. 
Subsequently, we also solved for amplitude time variation by gradually reducing the solution interval to 120, 60, and 30\,minutes. At each step, we confirmed that the coherence of the gain solutions was maintained.
As a result, J0108$-$2157 shows point-like structures and core-jet morphology elongated toward the west at 2.3 and 8.4\,GHz, respectively.
Its peak and integrated flux densities at 2.3\,GHz are 
151.5\,mJy\,beam$^{-1}$ and 170.9\,mJy, respectively, whereas at 8.4\,GHz, they are 70.1\,mJy\,beam$^{-1}$ and 87.4\,mJy, respectively.
To derive antenna-based gain corrections for the RR and LL visibilities separately \citep{EVNmemo78}, we applied the AIPS task CALIB for J0108$-$2157.
These gain solutions and the delay and rate solutions obtained through the fringe fitting were transferred to the target using the AIPS task CLCAL. 

To identify the target, whose location within the telescope's field of view was unknown, we initially utilized the task tclean in the Common Astronomy Software Applications \citep[CASA;][]{2007ASPC..376..127M} package, equipped with wide-field imaging mode and multiterm multi-frequency synthesis capabilities. 
This approach successfully detected the target at 8.4\,GHz, positioned $\sim 1$\,arcsecond from the phase center, but not at 2.3\,GHz. 
After adjusting the phase center of the visibility data to the location of the detected radio source using the AIPS task UVFIX, we conducted the subsequent imaging process for the 8.4-GHz data using the difmap software.
After integrating the visibilities every 30\,seconds, we made several rounds of CLEAN and phase-only self-calibration by gradually reducing the solution time interval to 120 and 60\,minutes.
Thus, we obtained the final image of the target at 8.4\,GHz.
Throughout the observations at the 8.4-GHz band, the amplitude fluctuations estimated by self-calibration to the phase calibrator consistently aligned with the absolute flux density uncertainties of 5\% for VLBA at 15\,GHz \citep{2002ApJ...568...99H}, which has also been applied to 8.4-GHz data \citep[e.g.,][]{2020ApJ...891...59R}.


\section{Results} \label{sec:res}
We detected a point source at R.A. = 01$^{\mathrm h}$\,02$^{\mathrm m}$\,49$^{\mathrm s}$.9911330\,$\pm\,0^{\mathrm s}.0000014$ and
DEC. = $-$22$^\circ$\,21$'$\,57.27013\,$\pm\,0.00042''$.
See Appendix\,\ref{sec:position} for estimates of the position errors (0.19 and 0.42\,mas in R.A. and DEC. directions, respectively).
The peak flux density in the naturally weighted image at 8.4\,GHz, which is optimized for sensitivity, exceeds the 10$\sigma$ level of the thermal noise.
The position of the target coincides with a continuum source detected at 239.4\,GHz \citep{2019ApJS..241...19I} and an X-ray point source observed by the Chandra X-ray Observatory \citep{2005ApJ...633..664T}.
Therefore, the radio source is likely to be associated with these submillimeter and X-ray sources.

\begin{figure*}[ht!]
    \plotone{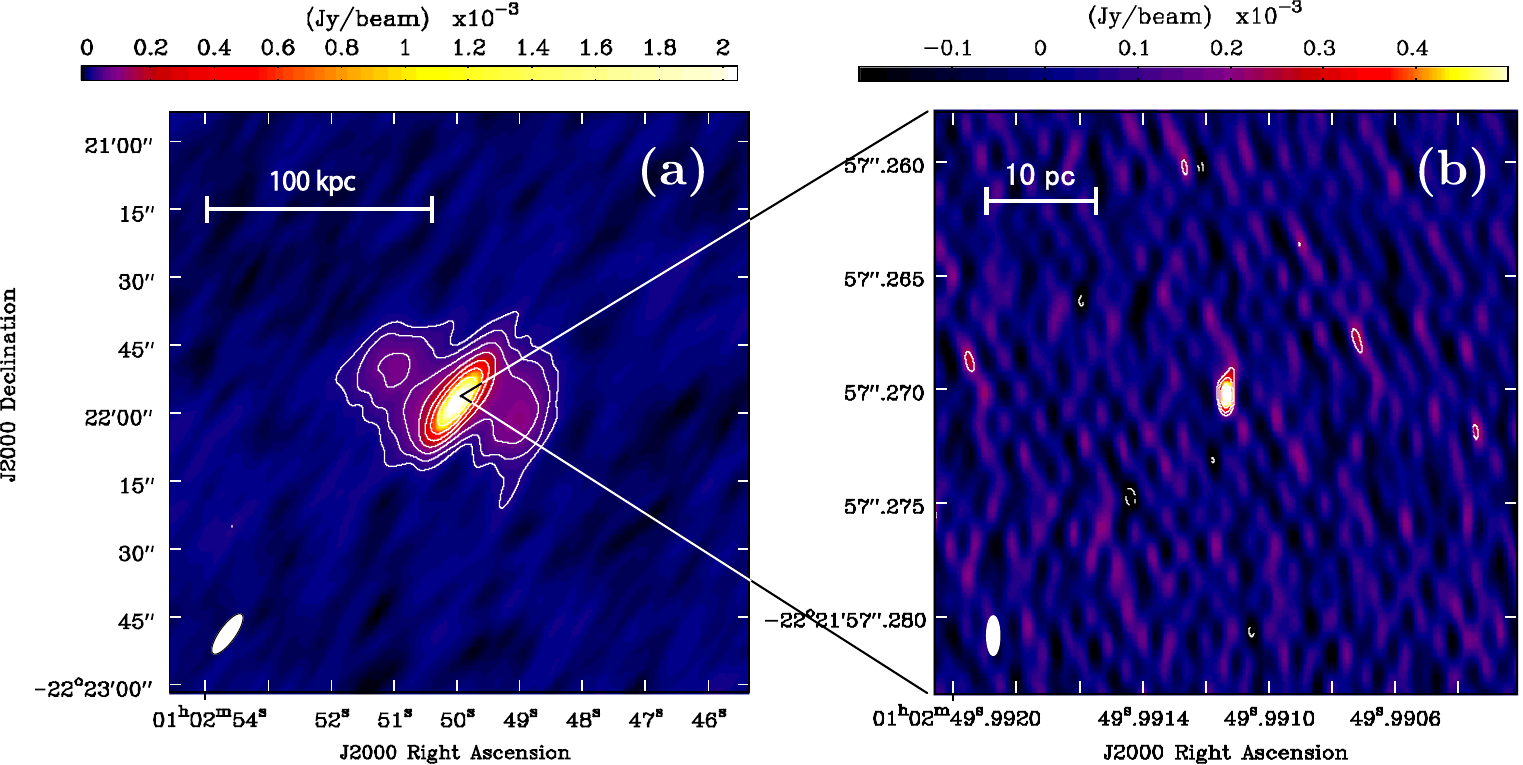}
    \caption{
    (a) A JVLA image at 9.0\,GHz with a resolution of $13.0 \times 4.7$\,arcsecond$^2$ at a PA of 143\,degrees, presented in \cite{2021MNRAS.504.2675H}. 
    Contour levels commence at 17\,$\mu$Jy\,beam$^{-1}$, corresponding to the 3$\sigma$ noise level of the image, 
    and are subsequently multiplied by a factor of 2. 
    The peak flux density in the image is 2.7\,mJy\,beam$^{-1}$.
    For more details on imaging processes, please see the reference paper.
    (b) A uniformly weighted VLBA image at 8.4\,GHz with a resolution of $1.75 \times 0.59$\,mas$^2$ at a PA of 0.1\,degrees. 
    Contour levels commence at 219\,$\mu$Jy\,beam$^{-1}$, corresponding to the 3$\sigma$ noise level of the image, 
    and are subsequently multiplied by a factor of $\sqrt{2}$. 
    The peak flux density in the image is 603\,$\mu$Jy\,beam$^{-1}$.
    \label{fig:X_image}}
\end{figure*}

Figure~\ref{fig:X_image} shows a uniformly weighted image of the target at 8.4\,GHz after phase-only self-calibration. 
The thermal noise and peak flux density measured in the image are 73 and 603\,$\mu$Jy\,beam$^{-1}$, respectively.
The Gaussian fit for the image performed using the AIPS task IMFIT yields a full-width half maximum (FWHM) of the source as $1.83\pm 0.23 \times 0.604 \pm  0.076$\,mas$^2$ at a position angle (PA) of $177.6 \pm 3.8$\,degrees on the image, which is similar to the beam size of the image. 
The IMFIT was unable to determine the deconvolved size. 
The only value obtained was that of the major axis, which was $\sim$0.5\,arcseconds.
We estimated the flux density of the source by summing the CLEAN components within the beam, yielding a value of 556\,$\mu$Jy\footnote{
The difmap software computes the peak flux density by converting the flux density per pixel in the image into flux density per beam size. 
Consequently, the flux density value of an unresolved source, smaller than the beam size, can be less than its peak flux density value.
}.
Based on these constraints, the brightness temperature of the source, $T_\mathrm{b}$, can be calculated using the equation:
\begin{eqnarray}
\left( \frac{T_{\mathrm b}}{\mathrm{K}} \right) =   \nonumber \\
1.8\times 10^{9} (1+z)
\left( \frac{S_\nu}{\mathrm{mJy}} \right)
\left( \frac{\nu}{\mathrm{GHz}} \right)^{-2}
\left( \frac{\phi}{\mathrm{mas}} \right)^{-2}
\label{eq:Tb},
\end{eqnarray}
where $\phi$ represents the source size \citep[cf.][]{2005ApJ...621..123U}.
Consequently, we obtained $T_\mathrm{b}\sim10^{7.2}$\,K for the detected source.
Nevertheless this value depends on the beam size of the observation.
We also measured the intrinsic source size by directly fitting a Gaussian component to the visibilities using the task modelfit in the difmap software and estimated the intrinsic brightness temperature, $T_\mathrm{b}'$.
The best-fit model is a Gaussian component with the integrated flux density of 561\,$\mu$Jy and deconvolved source size of $0.842 \times 0.168$\,mas$^2$ at a PA of $-14.7$\,degrees\footnote{The task modelfit in the difmap software does not output fitting errors.}, corresponding to the linear size of $\sim$1\,pc.
As a result, we obtained $T_\mathrm{b}'\sim 10^{8.1}$\,K.

\begin{figure}[ht!]
    \plotone{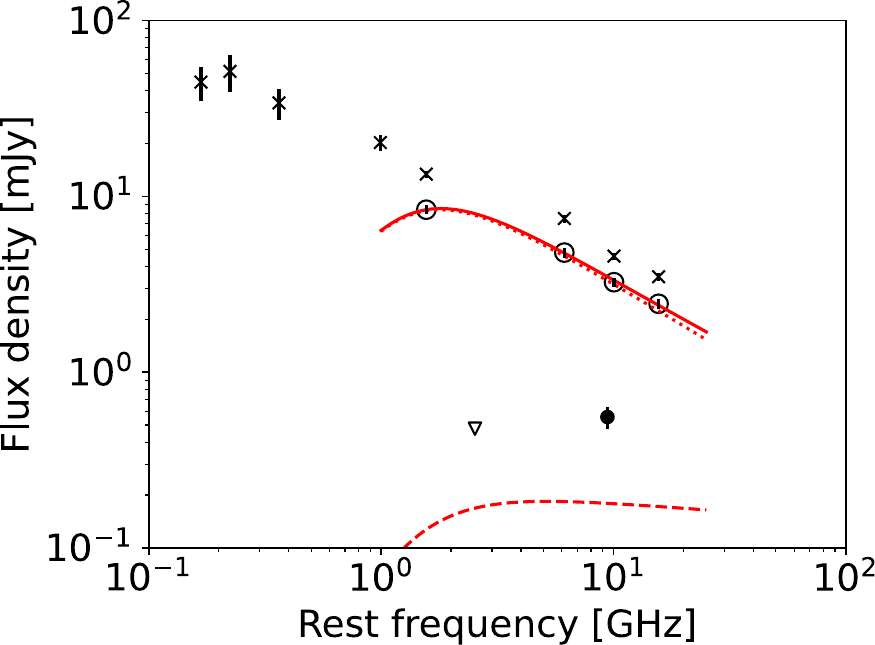}
    \caption{
    Radio spectrum of IRAS\,F01004$-$2237. 
    Crosses denote the integrated flux density of the source as provided by 
    TGSS at 150\,MHz \citep{2017A&A...598A..78I},
    GLEAM survey at 200\,MHz \citep{2017MNRAS.464.1146H},
    WISH at 352\,MHz \citep{2002A&A...394...59D},
    RACS at 887.5\,MHz \citep{2021PASA...38...58H},    
    and VLA observations by \cite{2021MNRAS.504.2675H}.
    Open circles represent the core flux density reported by \cite{2021MNRAS.504.2675H}.
    A filled circle and an open triangle depict the flux density at 8.4\,GHz and its 3$\sigma$ upper limit at 2.3\,GHz, respectively, obtained through VLBA observations in this study. 
    A red solid line illustrates the model for the VLA core as reported in \cite{2021MNRAS.504.2675H}, assuming a non-thermal component (dotted line) absorbed by ambient thermal plasma with free-free emission (dashed line).
    \label{fig:spectrum}
    }
\end{figure}

At 2.3\,GHz, we obtained a thermal noise level of 159\,$\mu$Jy\,beam$^{-1}$ for a uniformly weighted image with a beam size of $7.01 \times 2.27$\,mas$^2$ at a PA of $-1.6$\,degrees.
We adopted 3$\sigma$ as an upper limit of the flux density at the frequency.
As a result, the spectral index between 2.3 and 8.4\,GHz, $\alpha^{8.4}_{2.3}$, can be constrained as $\alpha^{8.4}_{2.3}>0.12$.
Figure \ref{fig:spectrum} illustrates the radio spectrum of the source with measurements for the extended structure obtained from previous studies.
Table~\ref{tbl:flux} summarizes our measurements on the images.

\begin{deluxetable*}{ccccccc}
\tabletypesize{\scriptsize}
\tablewidth{0pt} 
\tablecaption{Flux measurement of IRAS\,F01004$-$2237 \label{tbl:flux}}
\tablehead{										
	 \colhead{}	&	\colhead{}	&	\multicolumn{2}{c}{Source Size on the Image}			&	\colhead{} &	\colhead{} 	&	\colhead{} 	
  \\\cline{3-4}		
	 \colhead{Band} 	&	\colhead{RMS}   &	\colhead{FWHM} 	&	\colhead{PA} 	 &   \colhead{Peak flux density} 	&	\colhead{Integrated flux density} &   \colhead{Brightness Temperature} 	
  \\
	 \colhead{(GHz)} 	&	\colhead{($\mu$Jy\,beam$^{-1}$)} 	&	\colhead{(mas$^2$)} 	&	\colhead{(deg)} 	&  \colhead{($\mu$Jy\,beam$^{-1}$)} 	&	\colhead{($\mu$Jy)}  &   \colhead{(K)}
  \\
	 } 									
\startdata 										
	2.3	&	159 &	…	&	…	&    $<473$	&	$<473$	&   $<10^{7.1}$\\
	8.4	&	\phn73 &	$1.83\pm 0.23 \times 0.604 \pm  0.076$	&	 $177.6 \pm 3.8$	&   $603\pm79$	&	 $556\pm78$	&   $10^{7.2}$\\
\enddata    
\tablecomments{All values are measured in the uniformly weighted images. The flux density error is determined by taking the root mean square (RMS) of the thermal noise and the systematic uncertainty of amplitude calibration, set at 5\%.
}
\end{deluxetable*}

\section{Discussion} \label{sec:discuss}
\subsection{Origin of the Radio Emission}
The radio emission from the compact source in IRAS\,F01004$-$2237 has a high intrinsic brightness temperature of $T_\mathrm{b}'\sim10^{8.1}$\,K, as measured by the VLBA observation at 8.4\,GHz.
This feature indicates the presence of non-thermal phenomena, such as an AGN, RSN, or SNR, because compact starbursts that lack these  phenomena  have $T_{\mathrm b}' \lesssim 10^{5}$\,K \citep[][]{1992ARA&A..30..575C}.
Additionally, the source size of $\sim1$\,pc rules out the possibility of a single RSN or SNR.
The surface brightness and size of RSNe and SNRs typically exhibit a correlation, commonly known as the $\Sigma$-$D$ relation \citep{1976MNRAS.174..267C,1985ApJ...295L..13H}. 
According to this empirical relation, a 1-pc RSN or SNR has a surface brightness of $\sim10^{-15}$\,W\,m$^{-2}$\,Hz$^{-1}$\,str$^{-1}$ \citep[e.g.,][]{2005A&A...435..437U,2010A&A...509A..34B}, which corresponds to $T_{\mathrm b}'\sim10^{6}$\,K.
Although the presence of multiple compact and young SNRs/RSNe densely packed in the small area of $\sim$1\,pc might account for the observational results, such a scenario is deemed highly improbable considering the typical separation between RSNe/SNRs of $\sim$10\,pc in the nucleus of Arp\,220 \citep[e.g.,][]{2006ApJ...647..185L}.
Therefore, achieving $T_\mathrm{b}'\sim10^{8.1}$\,K is only possible through AGN activity.
This active radio source is a candidate for the AGN core that powers the 100-kpc extended radio emission reported by \cite{2021MNRAS.504.2675H}.

IRAS\,F01004$-$2237 exhibits a far-infrared to radio ratio of $q=2.49\pm0.04$ computed from the far-infrared flux density and the radio flux density at 1.4\,GHz \citep{2021MNRAS.504.2675H}, which aligns with the average value of 2.34 observed for entire galaxies \citep{2001ApJ...554..803Y}.
While low $q$ values, indicative of high radio power, are typically attributed to an AGN activity, IRAS\,F01004$-$2237, which hosts the AGN, does not exhibit such a characteristic.
Recent studies have revealed the presence of AGNs even in objects with moderate $q$ values \citep[e.g.,][]{2010ApJ...724..779M,2019PASJ...71...28S}, with IRAS\,F01004$-$2237 serving as an example of such an object.

Caution should be exercised when estimating star formation rates for these sources. 
According to the relation derived by \cite{2003A&A...409...99P}, a star formation rate is calculated as $258.4 \times 10^{-30} L_\mathrm{8.4}M_\odot$\,yr$^{-1}$, where $L_\mathrm{8.4}$ represents a specific radio luminosity at 8.4\,GHz in units of \ergs\,Hz$^{-1}$ \citep{2010MNRAS.405..887C}.
For IRAS\,F01004$-$2237, the star formation rate is obtained as $\sim300M_\odot$\,yr$^{-1}$ in the core region of the JVLA image \citep[Figure\,\ref{fig:X_image}a;][]{2021MNRAS.504.2675H}, assuming that all radio emission is associated with star formation \cite[e.g.,][]{2010MNRAS.405..887C}. 
However, we have found that AGN activity contributes to the radio emission of IRAS\,F01004$-$2237.
Therefore, at least the flux contribution from the compact radio source must be subtracted, resulting in a star formation rate of $\lesssim 250M_\odot$\,yr$^{-1}$. 
If the emission not detected by the VLBA observations is also of AGN origin, this constraint may even be an overestimate.

\subsection{Physical Constraints of the Active Radio Source}
The compact radio source originating from an AGN activity exhibits a spectral index of $\alpha^{8.4}_{2.3}>0.12$, suggesting the potential influence of Synchrotron self-absorption (SSA) and/or free-free absorption (FFA). 
Below, we discuss the possibility of each absorption mechanism in explaining the observed outcomes.

\begin{figure}[ht!]
    \plotone{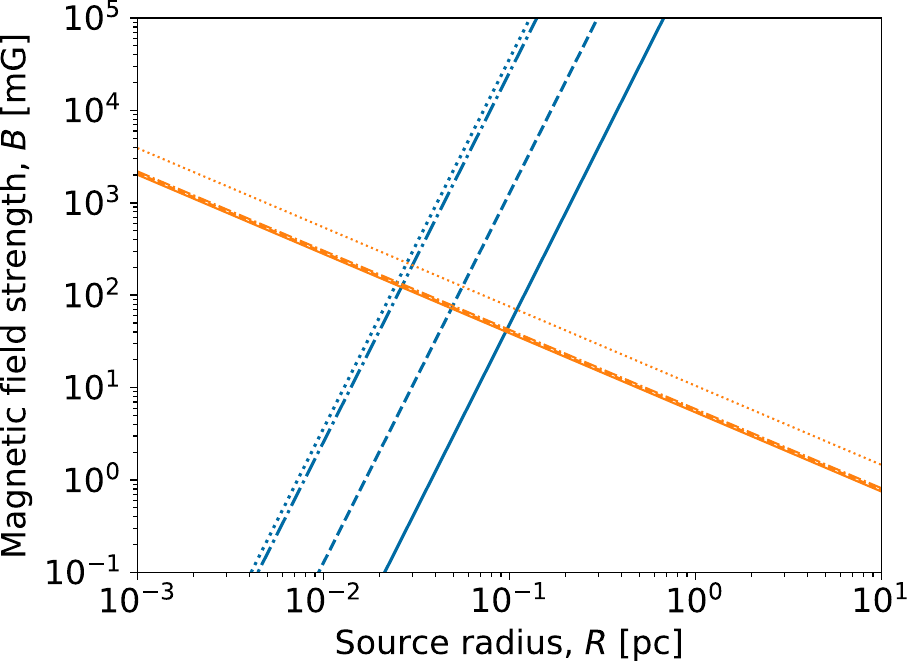}
    \caption{
    Physical conditions of the non-thermal plasma under the assumptions of SSA and equipartition. 
    The plot demonstrates the relationship between plasma radius, $R$, and magnetic field strength, $B$.
    The blue lines represent constraints imposed by SSA (Equation\,\ref{eq:B_SSA}), while the orange lines indicate constraints under the equipartition condition (Equation\,\ref{eq:B_eq}).
    The solid, dashed, dash-dotted, and dotted lines of each color correspond to peak frequencies at rest, $\nu_\mathrm{p}$, of 3, 5, 8, and 15\,GHz, respectively.  
    The equipartition condition given in Equation\,\ref{eq:B_eq} is insensitive to luminosity, resulting in the overlapping of the orange solid line with the dashed and dot-dashed lines.
    }
    \label{fig:SSA-Beq}
\end{figure}

The magnetic field strength responsible for SSA, $B_\mathrm{SSA}$, is determined by 
\begin{eqnarray}
\left( \frac{B_\mathrm{SSA}}{\mathrm{mG}} \right) =  \nonumber\\
3.1\times10^{4}
\left( \frac{\nu_{\mathrm p}}{\mathrm{GHz}} \right)^5
\left( \frac{S_\mathrm{p}}{\mathrm{mJy}} \right)^{-2} 
\left( \frac{\phi}{\mathrm{mas}} \right)^4
(1+z)^{-1}
\label{eq:B_SSA},
\end{eqnarray} 
where $S_{\mathrm p}$ is the peak flux density observed at frequency, $\nu_{\mathrm p}$ \citep{1981ARA&A..19..373K}.
To relate $S_{\mathrm p}$ to the observed flux density at 8.4\,GHz, we assume the spectral indices of the optically thin and thick parts to be $\alpha = -0.7$ and $2.5$, respectively.
On a different aspect, employing minimum energy conditions, where the energy densities of electrons and magnetic fields are roughly equal, we can compute the equipartition magnetic field strength, $B_{\mathrm eq}$, as
\begin{eqnarray}
\left( \frac{B_{\mathrm eq}}{\mathrm{mG}} \right) =
2.9\times10^{-11}
\left( \frac{R}{\mathrm{pc}} \right)^{-\frac{6}{7}}
\left( \frac{L}{\mathrm{erg/s}} \right)^{\frac{2}{7}}
(1+k)^{\frac{2}{7}}
\label{eq:B_eq}
,
\end{eqnarray}
where $R$, $L$, and $k$ represent 
the radius of the component, bolometric luminosity, and energy ratio between heavy particles and electrons \citep{1970ranp.book.....P}. 
We assume $k = 1$, but it is worth noting that $B_{\mathrm eq}$ remains relatively insensitive to the specific values of $k$.
We approximate $L \simeq 4\pi D_{\mathrm L}^2 S_{\mathrm p}\nu_{\mathrm p}$ for the bolometric luminosity \citep[e.g.,][]{2006A&A...456...97F}. 
Figure \ref{fig:SSA-Beq} represents the physical conditions constrained by Equations~(\ref{eq:B_SSA}) and (\ref{eq:B_eq}) for specific values of $\nu_\mathrm{p}$.




Assuming that SSA is a consequence of an equipartition magnetic field, i.e., $B_\mathrm{SSA}=B_{\mathrm eq}$, 
and the spectral peak is present at a few GHz, 
we derive a magnetic field strength on an order of $100$\,mG and a source size of an order of 0.01\,pc (see Figure \ref{fig:SSA-Beq}).
Although these values reasonably align with expectations for the base of the jet \citep[cf.][]{2023A&A...673A.159R}, we have obtained the deconvolved source size of an order of 1\,pc, which is inconsistent with the estimation assuming SSA. 
Therefore, SSA alone is not capable of explaining the physical condition of the active radio source.

\begin{figure}[ht!]
    \plotone{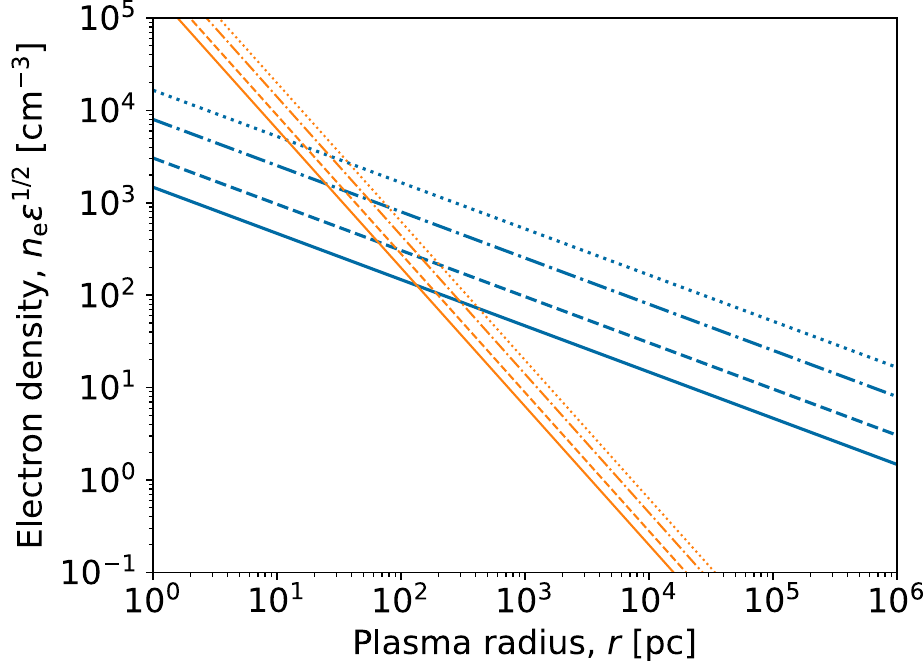}
    \caption{
    Physical conditions of the ambient thermal plasma assuming an electron temperature of $10^4$\,K.
    The plot demonstrates the relationship between the plasma radius, $r$, and electron density considering the filling factor, $n_\mathrm{e}\epsilon^{1/2}$.
    The blue lines represent constraints imposed by FFA  (Equation\,\ref{eq:tau}), with the solid, dashed, dash-dotted, and dotted lines corresponding to peak frequencies at rest, $\nu_\mathrm{p}$, of 1, 2, 5, and 10\,GHz, respectively. 
    The orange lines indicate the constraints imposed by FFE (Equation\,\ref{eq:FFE}), with the solid, dashed, dash-dotted, and dotted lines corresponding to flux density at optically thin regime due to FFE of 0.1, 0.2, 0.5, and 1.0\,mJy, respectively.
    \label{fig:FFA-FFE}}
\end{figure}

On the other hand, FFA can alternatively  contribute to the observed spectrum of the active radio source.
The substantial presence of ambient plasma associated with the host galaxy plays a significant role in enhancing the optical depth of FFA.
We can calculate the optical depth of FFA, $\tau_\nu$, as
\begin{eqnarray}
\tau_\nu = 
0.46 
\left( \frac{r}{\mathrm{pc}} \right) 
\left( \frac{T_\mathrm{e}}{\mathrm{K}} \right)^{-\frac{3}{2}} 
\left( \frac{n_\mathrm{e}}{\mathrm{cm}^{-3}} \right)^{2}
\left( \frac{\nu}{\mathrm{GHz}} \right) ^{-2.1}
\epsilon,
\label{eq:tau}
\end{eqnarray}
where $r$, $T_\mathrm{e}$, $n_\mathrm{e}$, and $\epsilon$ represent the thermal plasma's radius, electron temperature, electron density, and volume filling factor, respectively \citep[e.g.,][]{2001PhDT.......K}.
The spectrum peaks at the frequency where $\tau_\nu\sim 1$ is satisfied.
Figure~\ref{fig:FFA-FFE} depicts the physical condition of the plasma constrained by Equation\,(\ref{eq:tau}), assuming $T_\mathrm{e} = 10^4$\,K and certain values of $\nu_\mathrm{p}$
At the same time, the thermal plasma, which gives rise to FFA, radiates free-free emission (FFE).
In an optically thin regime, we can derive the electron density of the thermal plasma producing FFE as
\begin{eqnarray}
\left( \frac{n_\mathrm{e}}{\mathrm{cm}^{-3}} \right) = \nonumber \\
3.6\times10^3 
\left( \frac{r}{\mathrm{pc}} \right)^{-\frac{3}{2}} 
\left( \frac{T_\mathrm{e}}{\mathrm{K}} \right)^{\frac{1}{4}} 
\left( \frac{S_\mathrm{ff}}{\mathrm{Jy}} \right)^{\frac{1}{2}} 
\left( \frac{D_\mathrm{L}}{\mathrm{Mpc}} \right) 
\epsilon^{-1/2},
\label{eq:FFE}
\end{eqnarray}
where $S_\mathrm{ff}$ is the flux density at an optically thin regime  \citep[e.g.,][]{2009ApJ...693.1696H}.
FFE contributes an order of 0.1\,mJy to IRAS\,F01004$-$2237 in an optically thin regime \citep[][see Figure~\ref{fig:spectrum}]{2019ApJS..241...19I,2021MNRAS.504.2675H}.
Figure~\ref{fig:FFA-FFE} also illustrates the physical parameters constrained based on the conditions imposed by the FFE and assuming certain values of $S_\mathrm{ff}$.

Assuming the spectral peak is present at a few GHz, we estimate $r$ and $n_\mathrm{e}\epsilon^{1/2}$ to be an order of $100$\,pc and $100$\,cm$^{-3}$, respectively (see Figure~\ref{fig:FFA-FFE}).
These estimates are consistent with the conditions found in local ULIRGs, whose typical size of thermal plasma exhibiting FFE is $\sim$\,1\,kpc \citep{2017ApJ...843..117B}, and typical electron densities are $\sim$\,100\,cm$^{-3}$ determined through optical emission-line ratios \citep{1999ApJ...522..113V}. 
Therefore, FFA resulting from thermal plasma within the merging system can be a primary cause of the flat or inverted radio spectrum observed in the active radio source.

\subsection{Multiwavelength Characteristics and Evolutionary Implications of the AGN}
IRAS\,F01004$-$2237 has an optical counterpart located encompassing the compact radio source, which is classified as a type-2 Seyfert AGN \citep{2010ApJ...709..884Y,2013MNRAS.432..138R} and  hosts a supermassive black hole with a mass of $2.5 \times 10^7 M_\odot$ \citep{2006ApJ...651..835D}.
This AGN is currently active, marked by an increase of luminosity and spectral change in optical and infrared wavelengths attributed to tidal disruption events, whereby stars undergo gravitational disruption as they closely approach the supermassive black holes \citep{2017ApJ...841L...8D,2017NatAs...1E..61T,2021MNRAS.504.4377T}.
At X-ray, a point source has been identified at the coordinates of
R.A. = 01$^{\mathrm h}$\,02$^{\mathrm m}$\,49$^{\mathrm s}$.99, 
DEC. = $-$22$^\circ$\,21$'$\,57.3$''$ obtained through the
Chandra X-ray Observatory \citep{2005ApJ...633..664T} using the Advanced CCD Imaging Spectrometer (ACIS)\footnote{The reference paper does not provide a position error, which would typically be on the order of 0.1\,arcseconds. This estimate is based on the assumption that the ACIS observations with 1-arcsecond spatial resolution detect a signal greater than 5$\sigma$.}.
The position of the compact radio source aligns with the X-ray source, whose spectrum is very soft and lacks direct AGN emission below 10\,keV, suggesting a Compton-thick circumnuclear environment \citep{2011MNRAS.415..619N}. 
This argument is consistent with the infrared spectra of IRAS\,F01004$-$2237, which show signs of AGN obscuration \citep{2007ApJS..171...72I,2008PASJ...60S.489I,2009ApJS..182..628V,2010MNRAS.405.2505N}.

Despite the optical luminosity of the source not meeting the criteria, the host galaxy of IRAS\,F01004$-$2237 possesses a nearly stellar radial profile similar to that of quasars lacking tidal tails or loops \citep{1998ApJ...492..116S}, indicating an old merger \citep{2002ApJS..143..315V}. \cite{2021MNRAS.504.2675H} have mentioned that the absence of extended X-ray features \citep{1999A&A...349..389V,2005ApJ...633..664T} 
hints at the potential AGN origin of the 100-kpc scale radio emission.
Moreover, this object exhibits an HCN-to-HCO$^+$ ($J = 3$--2) flux-density ratio higher than unity in the northwest-southeast direction within its central few 100\,pc \citep{2019ApJS..241...19I}, where mechanical heating attributed to an AGN jet is one considerable explanation \citep[cf.][]{2012A&A...537A..44A,2013PASJ...65..100I,2024MNRAS.528.3668H}.
Therefore, we can infer that this object is a ULIRG that was once shrouded by dust and actively forming stars but is presently undergoing a process of dust dispersion due to AGN activities, including those of jets \citep[cf.][]{2023MNRAS.520.5712S}.

\subsection{Contribution of the Jet Activities to AGN Feedback}
In the following analysis, we explore whether the jet power of IRAS\,01004$-$2237 contribute to AGN feedback.
Under the minimum energy condition (cf. Equation\,\ref{eq:B_eq}), the time-averaged kinetic power of AGN jets fueling the extended features in a radio galaxy is derived as
\begin{eqnarray}
\left( \frac{P_\mathrm{jet}}{\mathrm{erg/s}} \right) 
= 3\times10^{15}
\left( \frac{L_{151}}{\mathrm{erg/s/Hz/sr}} \right)^{\frac{6}{7}}
f^{\frac{3}{2}},
\label{eq:Pkin_W99}
\end{eqnarray}
where $L_{151}$ ($=S_{151} D_\mathrm{L}^2$) represents the specific radio luminosity at 151\,MHz calculated from the flux density at 151\,MHz, $S_{151}$, and $f$ denotes a parameter accounting for systematic error in the model assumptions \citep{1999MNRAS.309.1017W}.
This relationship applies regardless of the radio power of jet activities \citep{2013ApJ...767...12G}.
Taking the flux density of $44.7\pm9.9$\,mJy provided by the TIFR GMRT Sky Survey \citep[TGSS;][]{2017A&A...598A..78I} at 150\,MHz as $S_{151}$, the specific radio luminosity of IRAS\,F01004$-$2237 is $L_{151}=(1.27\pm0.28)\times 10^{30}$\,\ergs\,Hz$^{-1}$\,sr$^{-1}$.
In this luminosity range, a value of $f=20$ (i.e., low radiative efficiency) can be applied \citep{2013ApJ...767...12G}.
Substituting these values to Equation\,(\ref{eq:Pkin_W99}), we obtain $P_\mathrm{jet} \sim 2\times10^{43}$\,\ergs.
Note that the spatial resolution of TGSS ($\sim$25\,arcseconds) is insufficient to separate the core from the extended structure as in Figure\,\ref{fig:X_image}, resulting in the measured flux density also including contributions from star formation. 
Therefore, the kinetic power of the jets estimated here represents an upper bound.

On another front, in IRAS\,F01004$-$2237, a blueshifted broad component with a $\mathrm{FWHM}\sim1400$\,\kms and a velocity shift of $\sim-800$\,\kms in the [\ion{O}{3}] emission line have been documented \citep{2013MNRAS.432..138R}.
Additionally, prominent blueshifted emission wings have been observed in the Ly$\alpha$ and \ion{N}{5} profiles \citep{2015ApJ...803....6M}, and a distinct P-Cygni profile has been detected in the OH doublet at 119\,$\mu$m \citep{2013ApJ...775..127S}, solidifying the presence of a thermal wind in this object. 
\cite{2018MNRAS.478.2438S} estimated its kinetic energy to be an order of $\sim 10^{41-42}$\,\ergs based on the [\ion{O}{3}] emission line.
Moreover, an \ion{O}{6} broad absorption line (BAL) is evident in IRAS\,F01004$-$2237, with a lower limit of its kinetic energy determined to be $10^{40.6}$\,\ergs\citep{2022ApJ...934..160L}.
Relative to the total AGN radiative power, which is $(8.8\pm0.4)\times 10^{44}$\,\ergs estimated from the [\ion{O}{3}] emission line, the contribution of the thermal outflow to the AGN feedback has been considered to be modest \citep{2018MNRAS.478.2438S}.

Although there is considerable uncertainty in estimating the transfer efficiencies of kinetic energies \citep[e.g.,][]{2012ApJ...757..136W,2017A&A...599A.123N}, 
the kinetic power of the jet in IRAS\,F01004$-$2237, $P_\mathrm{jet}$, can surpass that of the thermal outflow and  may possess sufficient potency to disperse the surrounding medium \citep[cf.][]{2013ApJ...779..173N}. 
Consequently, this ULIRG could be in a transition phase to become a quasar, accompanied by jet activity, as in the case observed in IRAS\,00182$-$7112 \citep{2007ApJ...654L..49S,2009ApJ...693.1223S,2012MNRAS.422.1453N}.
Further validation of this hypothesis will require the detection of jet emission at radio frequencies linking the pc-scale active radio source to the extended structure at the 100-kpc scale \citep[cf.][]{2010A&A...519L...5P,2023MNRAS.520.5712S}.
If this connection is confirmed, the involvement of jet activities in ULIRGs would emerge as a significant factor in galaxy evolution triggered by merger events.

\section{Summary} \label{sec:sum}
Our VLBA imaging of the ULIRG IRAS\,F01004$-$2237 has identified the pc-scale compact radio source with a intrinsic brightness temperature of  $T_{\mathrm b}'\sim10^{8.1}$\,K at 8.4\,GHz and a spectral index of $\alpha^{8.4}_{2.3}>0.12$. 
These characteristics indicate the presence of non-thermal phenomena, which cannot be attributed to a typical RSN or SNR, suggesting an AGN origin.
While the connection between the pc-scale and 100-kpc structures of the source still remains ambiguous, the identified active radio source is an AGN core potentially driving the extended features.
An investigation of the physical conditions has revealed that FFA primarily contributes to the observed radio spectrum.
Also from a multi-wavelength perspective, the AGN in IRAS\,F01004$-$2237 is obscured in a dense environment. 
The kinetic power of the jet, estimated from the 100-kpc extended features, can be greater than that of the thermal outflow, which could blow away the surrounding medium and contribute to AGN feedback during galaxy evolution.

\begin{acknowledgments}
This research has used the VizieR catalogue access tool, CDS, Strasbourg, France. 
Additionally, we utilized the NASA/IPAC Extragalactic Database (NED), operated by the Jet Propulsion Laboratory, California Institute of Technology, under contract with the National Aeronautics and Space Administration.
Cosmological calculations are conducted using a calculator provided by \cite{2006PASP..118.1711W}.
The NRAO operating VLBA is a facility of the National Science Foundation operated under a cooperative agreement by Associated Universities, Inc.
This work used the Swinburne University of Technology software correlator \citep{2011PASP..123..275D}, developed as part of the Australian Major National Research Facilities Programme and operated under license.
\end{acknowledgments}

\vspace{5mm}
\facilities{
    VLBA (NRAO)
    }
\software{
    AIPS \citep{2003ASSL..285..109G},
    astropy \citep{2013A&A...558A..33A,2018AJ....156..123A}
    CASA \citep{2007ASPC..376..127M}, 
    difmap \citep{1997ASPC..125...77S}. 
    }

\appendix
\section{Positional accuracy of the compact radio source in IRAS\,F01004$-$2237}\label{sec:position}
The position error of the compact radio source at 8.4\,GHz detected in IRAS\,F01004$-$2237 (0.19 and 0.42\,mas in R.A. and DEC. directions, respectively) was estimated by the root-sum-square of the individual astrometric error contributions \citep[e.g.,][]{2006A&A...452.1099P,2011Natur.477..185H}.
Table\,\ref{tbl:error} summarizes the estimated error budget in our phase-referencing observations.
The positional accuracy of the phase calibrator was obtained from the Radio Fundamental Catalog\footnote{http://astrogeo.org/sol/rfc/rfc\_2023d/}.
The flux peak accuracy for the target and calibrator was calculated by dividing the beam size by the signal-to-noise ratio of the uniformly weighted images.
We estimated that the residual errors arising from propagation delays due to the non-dispersive tropospheric medium \citep[e.g.,][]{1999ApJ...524..816R} and the dispersive ionospheric medium \citep[e.g.,][]{1990AJ.....99.1663L} by assuming a zenith angle of 65\,degrees and a zenith angle difference between the target and calibrator to be 1\,degree.
Based on the typical zenith delay used in the model for the VLBA correlator, we adopted a non-dispersive residual of 3\,cm \citep{1999ApJ...522..157R}.
The total electron content of the dispersive ionospheric medium was $\sim 2\times10^{17}$\,m$^{-2}$ over the sky of the VLBA stations, which was obtained from the US Total Electron Content Product Archive\footnote{https://www.ngdc.noaa.gov/stp/iono/ustec/index.html} maintained by the National Oceanic and Atmospheric Administration. 
The global ionospheric model based on GPS satellites has an accuracy of about 10--25\% \citep{1998RaSc...33..565M}. 
Conservatively, we adopted the uncertainty of 25\%, corresponding to $\sim 5\times10^{16}$\,m$^{-2}$.
The contributions due to Earth orientation parameters and antenna positions were estimated based on the simulation presented in \cite{2006A&A...452.1099P}.

\begin{deluxetable*}{ccccccccccc}
\tabletypesize{\scriptsize}
\tablewidth{0pt} 
\tablecaption{Error budget for the 8.4-GHz VLBA observation for IRAS\,F01004$-$2237 (in $\mu$as). \label{tbl:error}}
\tablehead{								
	 \colhead{Error component}	&	\colhead{R.A.}	&	\colhead{DEC.}}					
\startdata 								
	Source coordinates (calibrator) &	170   & 340  \\
	Flux peak accuracy (calibrator) &	\phn\phn2   & \phn\phn5 \\
	Flux peak accuracy (target) &	\phn72   & 210 \\
	Tropospheric residuals &	\phn45   & 126 \\
	Ionospheric residuals &	\phn\phn5   & \phn14 \\
    Antenna position & \phn\phn2 & \phn\phn8\\
    Earth orientation & \phn\phn1 & \phn\phn8\\
    Total & 190 & 419\\
\enddata								
\end{deluxetable*}

\bibliography{ULIRG_2023}{}
\bibliographystyle{aasjournal}



\end{CJK*}
\end{document}